# Comparison of models for dynamical structures of solar plasma

**Sergey N. Artekha**

Space Research Institute of RAS, Moscow 117997, Russia

E-mail: sergey.arteha@gmail.com (S.N. Artekha). ORCID: 0000-0002-4714-4277

**Abstract**

Possible mechanisms of generation of dynamical plasma structures are associated with the Schwarzschild criterion for convective instability and with excess energy (heat, magnetic): due to the superadiabatic temperature gradient, the overheated plasma starts to rise. In the framework of Navier-Stokes MHD equations, the complete system of equations for the full axisymmetric plasma model includes equations for the stream function, the azimuthal velocity component, the density change, the flow function of poloidal magnetic field, and the azimuthal component of the magnetic field. Within the framework of ideal hydrodynamics, the simplest plasma model is applicable to the initial stage of generation of plasma structures. In this case, rigorous analytical solutions are constructed from blocks. Within the framework of ideal hydrodynamics, some intermediate model – the refined model considers rotating plasma at the growth stage. In this case, some unidirectional differential rotation occurs. The full model studies the development of structures as rotating plasma from the very beginning to the highest development stage. The velocity field is adjusted to boundary conditions. Velocity components change in a coordinated manner. The stationary axisymmetric ambient magnetic field and its gradients have little effect on the plasma structure and velocity field. The movement structures for all models are similar. Radial flows converge towards the axis at the bottom, and there is an outflow to the periphery near the top of the plasma structure. The maximum vertical velocity component is reached at the center of the axis. Applications of these models to solar structures, including jets, are discussed.

## 1 Introduction

In laboratory and space plasma, both quasi-stationary and rapidly changing structures, including vortex structures, often arise. For example, convective plasma structures are known for the solar photosphere, such as granules (diameter $D\sim1500$ km, vertical convective flow $v_z\sim7$ km/s, lifetime $\tau \sim 20$ min); spicules ($D \sim 300$ km, $v_z \sim 15 \div 110$ km/s, $\tau \leq 15$ min) rise through the chromosphere, extending up to 10000 km; prominences (filaments) are also observed in the transition region and the solar corona, etc. Among plasma structures, special attention is drawn to suddenly emerging directed flows – jets. These are non-stationary, rapidly spreading elongated structures, limited in space, with a spiral movement of plasma inside. The speed of the plasma clot and the magnetic field reach maximum values within the structure and sharply decrease on its periphery. Jets are studied in controlled laboratory experiments (Wolff, Foest, Kersten, 2023; Safari and Sohbatzadeh, 2015). However, a wider variety of them is observed in uncontrolled natural phenomena: magnetospheric jets (Erickson and Wolf, 1980; Pontius and Wolf, 1990; Chen and Wolf, 1993; Grigorenko et al., 2014), solar coronal jets and loops (Scullion et al., 2009; Wedemeyer-Böhm et al., 2012; Raouafi et al., 2016; Cheung et al., 2015), and astrophysical jets (Blandford and Payne, 1982; Lovelace et al., 1986; Ferrari, 1998; Marshall et al., 2002). In nature, the characteristics of jets vary very widely (Raouafi et al., 2016; Shibata and Magara, 2011; Liu, Wang, Erdélyi, 2019; Tian et al., 2021). For example, for the Sun, some typical characteristics are: the transverse dimensions of the jets $r_0 = 1000 \div 5000$ km, length $L = 5000 \div 500000$ km and more, temperature $T = 10000 \div 4000000$ K, equilibrium temperature gradient $|\nabla T| = 1 \div 10$ K km$^{-1}$ (nonequilibrium one can be much larger), magnetic flux density $B = 20 \div 1000$ G, lifetime $\tau = 10 \div 90$ min; at the base of the jet, plasma speeds $v_0 = 10 \div 30$ km s$^{-1}$. Although there are differences in terminology, jets can be roughly divided by flow speed into slow ones with $v_z \leq 50$ km s$^{-1}$, medium ones with $v_z \sim 200$ km s$^{-1}$, and fast ones with $v_z \sim 1000$ km s$^{-1}$ (for comparison, for different solar layers from the photosphere to the lower part of the corona, the speed of sound is $c_s = 7 \div 150$ km s$^{-1}$, and the Alfvén speed is $v_A = 0.7 \div 3000$ km s$^{-1}$). We are interested in slow jets and jets with medium speeds (when $v_z < c_s$). Since the speeds of such jets are less than the orbital and escape velocities for the Sun ($v_I \approx 437$ km s$^{-1}$, $v_{II} \approx 618$ km s$^{-1}$), the flow slows down, gradually dissipates, and sooner or later returns back to the surface of the Sun (if this happens in a large surrounding volume, then the reverse flow of plasma will be at a low speed and almost imperceptible). This aspect will be discussed in detail in Section 7.

Thus, the study of dynamic plasma structures (both quasi-stationary and rapidly changing) is one of the fundamental problems of physics, representing great theoretical and practical interest. Although the theoretical description of axisymmetric plasma structures will be of a general character, for quantitative assessments we will consider the solar atmosphere.

Any dynamic plasma structure goes through several stages in its development: the emergence of an initial structure of flows (because of the creation of favorable boundary conditions for some plasma region), growth and achievement of the highest development (with maximum values of speed and fields), destruction of the structure (scattering or attenuation). We will be interested in the process of generation and in the stage of increasing flow velocities and magnetic fields.

Many works are devoted to the theoretical study of plasma jets, their structure, dynamics, mechanisms of generation and development (Chen and Wolf, 1999; Bogoyavlenskij, 2000; Onishchenko et al., 2015; Bellan, 2018; Onishchenko et al., 2018; Belyaev, Matafonov, and Zagreev, 2018; Onishchenko and Feygin, 2022). Despite many studies, there is still no generally accepted theory of plasma jets and existing models possess a few shortcomings (Bellan, 2018), for example, the solutions do not satisfy the conditions of decreasing at the periphery. Therefore, the search for the causes and mechanisms of generation and development of this phenomenon, the study of the jet's behavior in different conditions, and the construction of possible jet models are important tasks. The convective mechanism has been considered previously (Wedemeyer-Böhm et al., 2012; Bogoyavlenskij, 2000; Bellan, 2018; Onishchenko et al., 2021), but they provided approximate solutions only for a limited region of the jet. It is also necessary to mention the work on three-dimensional numerical MHD simulations of jet formation (Fan and Gibson, 2004; Zhu et al., 2023; Gonzalez-Aviles et al., 2020; Wyper, DeVore, Antiochos, 2019; Li et al., 2025). The work (Gonzalez-Aviles et al., 2020) performed simulations of different parameters (including the resistivity and thermal conductivity) to check the impacts on the origin and morphology of the jets. The work (Wyper, DeVore, Antiochos, 2019) simulated a helical jet caused by a sigmoidal flux rope eruption, with artificial surface motions injecting free energy; some of the most energetic jets occur on the periphery of active regions where the magnetic field is strongly inclined. Most of the simulations about jets are ideal MHD models controlled by artificial settings: "preexisting twisted magnetic flux tubes"; "adding an anomalous (artificial) resistivity" and "trigger – rotational motion of flux rope"; "preexisting two magnetic loops close together with opposite polarity"; artificially given enhancing "surface motions"; "unstable flux rope embedded in a fan-spine configuration" etc. Obviously, these are forced restrictions, since the mechanism and all nuances of formation of jets and the conditions of their occurrence are not yet precisely known.

We will consider the following models. The full axisymmetric jet model can describe the plasma structures within the framework of the Navier-Stokes MHD equations at any stage of their development. Within the framework of ideal hydrodynamics, the simplest analytical model for the initial stage of generation of non-relativistic jets was presented in (Onishchenko, Artekha, Artekha, 2024). This model is based on the concept of Schwarzschild convective instability. The refined model takes into account all nonlinear terms within the framework of ideal hydrodynamics and can describe rotating plasma structures at the growth stage.

The goals of this work are the following:

– within the framework of the simple analytical model with several free parameters (Onishchenko, Artekha, Artekha, 2024), to describe the structure of the jet velocity field and magnetic field for some nonlinear z-profiles;

– in the approximation of ideal MHD, to derive a refined system of equations and obtain their numerical solution for the description of an initial stage of development of plasma structures in a stratified plasma;

– considering dissipation, to obtain numerical solutions for the description of velocity and magnetic fields at any stage of generation and development of plasma structures;

– to compare the numerical results obtained for all three models.

The structure of the article is as follows. Section 2 presents the general system of MHD equations that can manage developing plasma structure. Section 3 considers various simplifications of the resulting system of equations in the dissipative case and for ideal MHD. Section 4 discusses the structure of the velocity field and magnetic field for linear and nonlinear z-profiles within the framework of the simplest convective model. Section 5 describes numerical solutions of the refined convective model. Section 6 presents numerical solutions of the complete system of equations with the transition to the highest development stage. Section 7 discusses the main findings of the study and contains conclusions.

## 2 System of basic equations

We will study the generation of plasma structures in a cylindrical coordinate system $(r, \varphi, z)$ with the $z$ axis directed vertically upward. In axisymmetric case, all quantities are independent of the angle of the cylindrical coordinate system $\varphi$, i.e. $\partial/\partial\varphi = 0$. The flow velocity is $\boldsymbol{v} = (v_r, v_\varphi, v_z)$. The values of pressure, density, and magnetic induction, respectively, can be written by the sum of the initial and perturbed values:

$$p = p_0(z) + \tilde{p}(t,r,z),\ \rho = \rho_0(z) + \tilde{\rho}(t,r,z),\ \mathbf{B} = \mathbf{B}_0 + \widetilde{\boldsymbol{B}}, \tag{1}$$

where $\boldsymbol{B}_0 = (B_{0r}, B_{0\varphi}, B_{0z})$; near the magnetic pole (or a local field output) $\boldsymbol{B}_0 = \boldsymbol{B}_{0z}$. At the initial moment, we have for the equilibrium state:

$$\frac{dp_0}{dz} = -\rho_0(z)g,\ \ \frac{d}{dr}\left(p_0 + \frac{B_0^2}{2\mu_0}\right) = 0, \tag{2}$$

where $\mu_0 = 1.256637 \cdot 10^{-6}\ \mathrm{N \cdot A^{-2}}$ is the magnetic permeability of vacuum, $\boldsymbol{g} = -g\boldsymbol{e}_z$ is the gravitational acceleration on the Sun (star, or in the magnetosphere of the planet), $\boldsymbol{e}_z$ is a unit vector along the vertical. Gas pressure prevails over magnetic pressure in the solar photosphere and chromosphere (the value of $\beta = 2\mu_0 p_0/B_0^2 \geq 1$), but in the solar corona the situation is opposite ($\beta = \frac{2\mu_0 p_0}{B_0^2} < 1$). Although we will not need an explicit expression for the equilibrium pressure in what follows, we will nevertheless give some approximations. For plasma structures of small length, when we can assume that $\Delta T \ll T$, that is, the plasma layer is practically isothermal, we obtain for the equilibrium pressure:

$$p_0(z) = p_{00}\, exp\left(-\frac{z}{H}\right),\ \ H = \frac{c_s^2}{\gamma_a g},\ \ c_s = \sqrt{\frac{\gamma_a p_{00}}{\rho_{00}}}, \tag{3}$$

where $H$ is the local scale of atmospheric height, $c_s$ is the speed of sound, $\gamma_a$ is the adiabatic exponent; $p_{00}$ and $\rho_{00}$, respectively, are the initial values of pressure and density at the bottom of the plasma structure (at $z = 0$). For structures of greater length, it is necessary to take into account the change in temperature with height $\frac{dT}{dz} = l_0$ ; then we obtain the following approximation for the equilibrium pressure:

$$p_0(z) = p_{00}\left(1 + \frac{l_0 z}{T_{00}}\right)^{-gM/(R_u l_0)}, \tag{3*}$$

here $g$ is the gravitational acceleration, $M$ is the molar mass, $R_u$ is the universal gas constant. The explicit expression for the equilibrium density $\rho_0(z)$, which we need for numerical calculations, will be given further in Section 5.

The general system of magnetohydrodynamic equations consists of the following equations (in the cylindrical coordinate system):

1) continuity equation:

$$\frac{\partial \rho}{\partial t} = -\nabla \cdot (\rho \boldsymbol{v}) = -\rho(\nabla \cdot \boldsymbol{v}) - (\boldsymbol{v} \cdot \nabla)\rho.$$

The conditions under which a fluid can be considered incompressible ($\rho = constant$) are well known (Landau and Lifshitz, 1987): $\Delta v \ll c_s, \tau \gg L/c_s$, where $\tau$ and $L$ are the characteristic time and distance at which the speed undergoes noticeable changes $\Delta v$. The condition of fluid incompressibility is a very strong condition. The condition of incompressibility of the flow ($\nabla \cdot \boldsymbol{v} = 0$) is much weaker. Since the development time of many plasma structures is $\tau > L/c_s$, we will consider the arising flow to be incompressible:

$$\frac{1}{r}\frac{\partial(rv_r)}{\partial r} + \frac{\partial v_z}{\partial z} = 0, \tag{4}$$

then the perturbed density $\tilde{\rho}(t, r, z)$ can be found from the equation:

$$\frac{\partial \tilde{\rho}}{\partial t} = -v_r \frac{\partial \tilde{\rho}}{\partial r} - v_z \frac{\partial(\rho_0 + \tilde{\rho})}{\partial z}. \tag{5}$$

Due to the above limitations, we will not consider fast coronal jets.

2) Navier-Stokes MHD equations:

$$\frac{\partial v_r}{\partial t} + v_r \frac{\partial v_r}{\partial r} + v_z \frac{\partial v_r}{\partial z} - \frac{v_\varphi^2}{r} = -\frac{1}{\rho}\frac{\partial}{\partial r}\left(p + \frac{B^2}{2\mu_0}\right) +$$

$$+\nu\left(\frac{1}{r}\frac{\partial}{\partial r}\left(r\frac{\partial v_r}{\partial r}\right) + \frac{\partial^2 v_r}{\partial z^2} - \frac{v_r}{r^2}\right) + \frac{1}{\mu_0 \rho}\left(B_r \frac{\partial B_r}{\partial r} + B_z \frac{\partial B_r}{\partial z} - \frac{B_\varphi^2}{r}\right), \tag{6}$$

$$\frac{\partial v_z}{\partial t} + v_r \frac{\partial v_z}{\partial r} + v_z \frac{\partial v_z}{\partial z} = -\frac{1}{\rho}\frac{\partial}{\partial z}\left(p + \frac{B^2}{2\mu_0}\right) +$$

$$+\nu\left(\frac{1}{r}\frac{\partial}{\partial r}\left(r\frac{\partial v_z}{\partial r}\right) + \frac{\partial^2 v_z}{\partial z^2}\right) + \frac{1}{\mu_0 \rho}\left(B_r \frac{\partial B_z}{\partial r} + B_z \frac{\partial B_z}{\partial z}\right), \tag{7}$$

$$\frac{\partial v_\varphi}{\partial t} + v_r \frac{\partial v_\varphi}{\partial r} + v_z \frac{\partial v_\varphi}{\partial z} + \frac{v_r v_\varphi}{r} = \nu\left(\frac{1}{r}\frac{\partial}{\partial r}\left(r\frac{\partial v_\varphi}{\partial r}\right) + \frac{\partial^2 v_\varphi}{\partial z^2} - \frac{v_\varphi}{r^2}\right) +$$

$$+\frac{1}{\mu_0 \rho}\left(B_r \frac{\partial B_\varphi}{\partial r} + B_z \frac{\partial B_\varphi}{\partial z} + \frac{B_\varphi B_r}{r}\right). \tag{8}$$

3) Equation of state of an ideal gas with the temperature *T*:

$$\frac{p}{\rho T} = const. \tag{9}$$

4) The transport equation for the potential temperature $\theta = p^{1/\gamma_a}/\rho$:

$$\frac{d\theta}{dt} = 0. \tag{10}$$

5) Gauss's law:

$$\frac{1}{r}\frac{\partial(rB_r)}{\partial r} + \frac{\partial B_z}{\partial z} = 0, \tag{11}$$

6) Equations for magnetic induction:

$$\frac{\partial B_r}{\partial t} + v_r \frac{\partial B_r}{\partial r} + v_z \frac{\partial B_r}{\partial z} - B_r \frac{\partial v_r}{\partial r} - B_z \frac{\partial v_r}{\partial z} = \frac{1}{\mu_0 \sigma}\left(\frac{1}{r}\frac{\partial}{\partial r}\left(r\frac{\partial B_r}{\partial r}\right) + \frac{\partial^2 B_r}{\partial z^2} - \frac{B_r}{r^2}\right), \tag{12}$$

$$\frac{\partial B_z}{\partial t} + v_r \frac{\partial B_z}{\partial r} + v_z \frac{\partial B_z}{\partial z} - B_r \frac{\partial v_z}{\partial r} - B_z \frac{\partial v_z}{\partial z} = \frac{1}{\mu_0 \sigma}\left(\frac{1}{r}\frac{\partial}{\partial r}\left(r\frac{\partial B_z}{\partial r}\right) + \frac{\partial^2 B_z}{\partial z^2}\right), \tag{13}$$

$$\frac{\partial B_\varphi}{\partial t} + v_r \frac{\partial B_\varphi}{\partial r} + v_z \frac{\partial B_\varphi}{\partial z} + \frac{B_r v_\varphi}{r} - B_r \frac{\partial v_\varphi}{\partial r} - B_z \frac{\partial v_\varphi}{\partial z} - \frac{v_r B_\varphi}{r} = \frac{1}{\mu_0 \sigma}\left(\frac{1}{r}\frac{\partial}{\partial r}\left(r\frac{\partial B_\varphi}{\partial r}\right) + \frac{\partial^2 B_\varphi}{\partial z^2} - \frac{B_\varphi}{r^2}\right). \tag{14}$$

The formation of a jet can be caused by an excess of any locally released energy from the Sun (magnetic energy, heat energy etc.), which can be transformed into kinetic energy, or by convective instability (which can also act as a starting mechanism). It is important that the pushing force exceeds the restoring force. The instability of the stratified plasma leads to the generation of vertical flows. It is well known that the Brunt–Väisälä frequency characterizes internal gravity waves. Its square is (see local values from formula (3)):

$$\omega_g^2 \equiv \frac{g}{\theta}\frac{d\theta}{dz} = g\left(\frac{\gamma_a - 1}{\gamma_a H} + \frac{1}{T}\frac{dT}{dz}\right). \quad (15)$$

In the case of $\omega_g^2 > 0$, any disturbances are carried away from the region of their origin by the internal gravity waves, and the plasma remains to be stable. The plasma is unstable stratified if $\omega_g^2 < 0$, i.e. if the vertical temperature gradient $dT/dz$ is negative and its absolute value exceeds the first term in (15). The Schwarzschild criterion for convective instability is:

$$-\frac{1}{T}\frac{dT}{dz} > \rho_0 g \frac{\gamma_a - 1}{\gamma_a p_0}.$$

In such a case, disturbances can turn into unstable growing structures.

In Section 3 we will consider the problem starting from the most complete system of equations derived from the equations of this Section and ending with the simplest system of equations after all simplifications. When simplifying the basic system of equations, we will follow the procedure developed in (Onishchenko, Pokhotelov, Horton, 2015).

## 3 Derivation of simplified systems of equations for three models

To satisfy the equation (4), we will use the stream function in cylindrical coordinates. Radial and vertical velocity components are obtained from the stream function $\psi(t, r, \varphi, z)$ as follows:

$$v_r = -\frac{1}{r}\frac{\partial \psi}{\partial z}, \quad v_z = \frac{1}{r}\frac{\partial \psi}{\partial r}. \quad (16)$$

We will not write out equation (14) explicitly after these substitutions. The equation for density change (5) can be written in terms of the stream function and $\chi = \frac{\tilde{\rho}}{\rho_0}$ as follows:

$$r\frac{\partial \chi}{\partial t} + J(\psi, \chi) + (1+\chi)\frac{\partial \psi}{\partial r}\frac{d\, ln\, \rho_0}{dz} = 0, \quad (17)$$

where $J(a, b) = \frac{\partial a}{\partial r}\frac{\partial b}{\partial z} - \frac{\partial a}{\partial z}\frac{\partial b}{\partial r}$ is the Jacobian determinant.

To satisfy the equation (11), we will use the flow function for poloidal magnetic field $\psi_B(t, r, \varphi, z)$ (in cylindrical coordinates). Radial and vertical components of total magnetic flux density (from ambient and generated magnetic fields) are obtained as follows:

$$B_r = B_{0r} - \frac{1}{r}\frac{\partial \psi_B}{\partial z}, \quad B_z = B_{0z} + \frac{1}{r}\frac{\partial \psi_B}{\partial r}. \quad (18)$$

The equation (8) for the azimuthal velocity component can be written as follows:

$$\frac{\partial v_\varphi}{\partial t} - \frac{1}{r}\frac{\partial \psi}{\partial z}\frac{\partial v_\varphi}{\partial r} + \frac{1}{r}\frac{\partial \psi}{\partial r}\frac{\partial v_\varphi}{\partial z} - \frac{\partial \psi}{\partial z}\frac{v_\varphi}{r^2} = \nu\left(\frac{1}{r}\frac{\partial v_\varphi}{\partial r} + \frac{\partial^2 v_\varphi}{\partial r^2} + \frac{\partial^2 v_\varphi}{\partial z^2} - \frac{v_\varphi}{r^2}\right) +$$
$$+\frac{1}{\mu_0\rho}\left[\left(B_{0z} + \frac{1}{r}\frac{\partial \psi_B}{\partial r}\right)\frac{\partial B_\varphi}{\partial z} + \left(B_{0r} - \frac{1}{r}\frac{\partial \psi_B}{\partial z}\right)\frac{\partial B_\varphi}{\partial r} + \frac{1}{r}\left(B_{0r} - \frac{1}{r}\frac{\partial \psi_B}{\partial z}\right)B_\varphi\right]. \quad (19)$$

Using equations (6) and (7), we can exclude the total pressure gradient and find the $\varphi$-component of the curl of the velocity. As a result, we obtain the following equation:

$$\frac{\partial}{\partial t}\Delta^*\psi + \frac{1}{r}J(\psi, \Delta^*\psi) + \frac{2}{r^2}\frac{\partial \psi}{\partial z}\Delta^*\psi + 2v_\varphi\frac{\partial v_\varphi}{\partial z} + \nu r\Delta\left(\frac{1}{r}\Delta^*\psi\right) + \frac{\nu}{r^2}\frac{\partial^2 \psi}{\partial z^2} + gr\frac{\partial \ln\rho}{\partial r} +$$
$$+\frac{\partial \ln\rho}{\partial r}\left(\frac{\partial}{\partial t}\frac{\partial \psi}{\partial r} + \frac{1}{r}\frac{\partial \psi}{\partial r}\frac{\partial^2 \psi}{\partial z\partial r} - \frac{\partial \psi}{\partial z}\frac{\partial}{\partial r}\left(\frac{1}{r}\frac{\partial \psi}{\partial r}\right)\right) + \frac{\partial \ln\rho}{\partial z}\left(\frac{\partial}{\partial t}\frac{\partial \psi}{\partial z} + \frac{1}{r}\frac{\partial \psi}{\partial r}\frac{\partial^2 \psi}{\partial z^2} + v_\varphi^2 - \frac{\partial \psi}{\partial z}\frac{\partial}{\partial r}\left(\frac{1}{r}\frac{\partial \psi}{\partial z}\right)\right) +$$
$$+\frac{r}{\mu_0\rho}\frac{\partial}{\partial z}\left[\left(B_{0r} - \frac{1}{r}\frac{\partial \psi_B}{\partial z}\right)\frac{\partial}{\partial r}\left(B_{0r} - \frac{1}{r}\frac{\partial \psi_B}{\partial z}\right) + \left(B_{0z} + \frac{1}{r}\frac{\partial \psi_B}{\partial r}\right)\frac{\partial}{\partial z}\left(B_{0r} - \frac{1}{r}\frac{\partial \psi_B}{\partial z}\right) - B_\varphi^2\right] -$$
$$-\frac{r}{\mu_0\rho}\frac{\partial}{\partial r}\left[\left(B_{0r} - \frac{1}{r}\frac{\partial \psi_B}{\partial z}\right)\frac{\partial}{\partial r}\left(B_{0z} + \frac{1}{r}\frac{\partial \psi_B}{\partial r}\right) + \left(B_{0z} + \frac{1}{r}\frac{\partial \psi_B}{\partial r}\right)\frac{\partial}{\partial z}\left(B_{0z} + \frac{1}{r}\frac{\partial \psi_B}{\partial r}\right)\right] = 0, \quad (20)$$

where the Laplace operator $\Delta$ and the Grad–Shafranov operator $\Delta^*$, respectively, are defined as

$$\Delta = \frac{1}{r}\frac{\partial}{\partial r}\left(r\frac{\partial}{\partial r}\right) + \frac{\partial^2}{\partial z^2}, \quad (21)$$

$$\Delta^* = r\frac{\partial}{\partial r}\left(\frac{1}{r}\frac{\partial}{\partial r}\right) + \frac{\partial^2}{\partial z^2}. \quad (22)$$

All the equations obtained above are derived without any simplifications. From equations (12) and (13), we find an equation for the $\varphi$-component of the magnetic induction curl:

$$\frac{\partial}{\partial t}\Delta^*\psi_B - r\frac{\partial}{\partial t}\mathrm{curl}_\varphi\,\mathbf{B_0} + \frac{1}{r}J(\psi, \Delta^*\psi_B) - \frac{1}{r}J(\psi_B, \Delta^*\psi) + \frac{1}{\mu_0\sigma}\left(r\Delta\left(\frac{1}{r}\Delta^*\psi_B\right) + \frac{1}{r^2}\frac{\partial^2 \psi_B}{\partial z^2}\right) +$$
$$+\frac{\partial}{\partial z}\left[\frac{\partial \psi}{\partial z}\frac{\partial B_{0r}}{\partial r} - \frac{\partial \psi}{\partial r}\frac{\partial B_{0r}}{\partial z} - B_{0r}r\frac{\partial}{\partial r}\left(\frac{1}{r}\frac{\partial \psi}{\partial z}\right) - B_{0z}\frac{\partial^2 \psi}{\partial z^2} + \frac{1}{\mu_0\sigma}\left(\frac{\partial}{\partial r}\left(r\frac{\partial B_{0r}}{\partial r}\right) + r\frac{\partial^2 B_{0r}}{\partial z^2} - \frac{B_{0r}}{r}\right)\right] -$$
$$-r\frac{\partial}{\partial r}\left[\frac{1}{r}\frac{\partial \psi}{\partial z}\frac{\partial B_{0z}}{\partial r} - \frac{1}{r}\frac{\partial \psi}{\partial r}\frac{\partial B_{0z}}{\partial z} + B_{0r}\frac{\partial}{\partial r}\left(\frac{1}{r}\frac{\partial \psi}{\partial r}\right) + \frac{B_{0z}}{r}\frac{\partial^2 \psi}{\partial r\partial z} + \frac{1}{\mu_0\sigma}\left(\frac{1}{r}\frac{\partial}{\partial r}\left(r\frac{\partial B_{0z}}{\partial r}\right) + \frac{\partial^2 B_{0z}}{\partial z^2}\right)\right] +$$
$$+\frac{2}{r}\left(\Delta_r^*\psi\,\Delta_r^*\frac{\partial \psi_B}{\partial z} - \Delta_r^*\psi_B\,\Delta_r^*\frac{\partial \psi}{\partial z} + \frac{\partial^2 \psi}{\partial r\partial z}\frac{\partial^2 \psi_B}{\partial z^2} - \frac{\partial^2 \psi_B}{\partial r\partial z}\frac{\partial^2 \psi}{\partial z^2}\right) = 0. \quad (23)$$

Thus, the most complete system includes equations (14), (17), (19), (20) and (23). This system describes the full model of a dynamic plasma structure and can be solved by numerical methods only. The complete system of equations can determine the values of $\psi, v_\varphi, \rho, \psi_B, B_\varphi$ at arbitrary stage of development of the structure. Apparently, fields greater than 100 G are generated by dynamic structures of the jet itself.

Although the magnetic field always affects the velocity field through magnetic pressure and magnetic tension, nevertheless, as shown in the work (Onishchenko, Artekha, Artekha, 2024), the influence of the magnetic field on the process of generating plasma structures in the region with $\mathbf{B} = (0,0,B_z)$ is insignificant at the initial stage. At $t < 0$, the plasma is in dynamic equilibrium (including taking into account magnetic pressure). At $t = 0$, instability occurs, meaning that all additions to the physical quantities should be considered as small perturbations (including density changes and generated fields). A standard linearization procedure can be performed for all these additions (to describe the linear stage of generation). If we substitute a uniform external magnetic field $B_{z0}$ into the full system of the magnetohydrodynamic equations in a cylindrical coordinate system, it is easy to see that, to a first approximation (at the generation stage), the magnetic field has a weak effect on the velocity field. This effect can become noticeable only if the generated nonuniform magnetic field increases significantly (as a secondary effect at nonlinear stage). We are interested in the $\varphi$-components of vorticity and velocity. So, if the vertical component of the magnetic field prevails on the scale of the structure and the field changes $\frac{L}{B_{0z}}\frac{\partial \tilde{B}_z}{\partial z} \ll 1$ and $\frac{R_0}{B_{0z}}\frac{\partial \tilde{B}_z}{\partial r} \ll 1$ are small, then the only magnetic term $-\frac{1}{\mu_0}\frac{\partial}{\partial r}\left(\frac{B_{0z}}{\rho}\frac{\partial \tilde{B}_z}{\partial z}\right)\boldsymbol{i}_\varphi$ is negligible. For large fast coronal jets and loops (which we do not consider), these conditions are often violated. However, for slow spicules, small and medium jets, these conditions are often satisfied. In this case, the main mechanism of generation of plasma structures can be the convective mechanism (at the initial stage). Of course, strong magnetic field always creates a preferential direction of propagation along the magnetic field (as a result, $v_z \gg v_\varphi \gg v_r$ , $B_z \geq B_\varphi \gg B_r$). Within the framework of ideal MHD, neglecting small nonlinear terms, we can write a simpler system of equations describing the quantities $\psi, v_\varphi, \rho$ at the stage of intensification of the structure. Instead of (17), (19) and (20), respectively, we have the following equations:

$$r\frac{\partial \tilde{\chi}}{\partial t} + J(\psi, \tilde{\chi}) - \omega_g^2 \frac{\partial \psi}{\partial r} = 0, \tag{24}$$

$$\frac{\partial v_\varphi}{\partial t} - \frac{1}{r}\frac{\partial \psi}{\partial z}\frac{\partial v_\varphi}{\partial r} + \frac{1}{r}\frac{\partial \psi}{\partial r}\frac{\partial v_\varphi}{\partial z} - \frac{\partial \psi}{\partial z}\frac{v_\varphi}{r^2} = 0, \tag{25}$$

$$\left(\frac{\partial^2}{\partial t^2} + \omega_g^2\right)\Delta^*\psi + \frac{d\ln\rho_0}{dz}\frac{\partial}{\partial t}\left(\frac{\partial^2\psi}{\partial z\partial t} + v_\varphi^2\right) + \frac{\partial}{\partial t}\left(\frac{1}{r}J(\psi, \Delta^*\psi) + \frac{2}{r^2}\frac{\partial \psi}{\partial z}\Delta^*\psi + 2v_\varphi\frac{\partial v_\varphi}{\partial z}\right) = 0, \tag{26}$$

where $\tilde{\chi} = g\frac{\tilde{\rho}}{\rho_0}$. The simpler system of equation (24)-(26) describes some intermediate model and can also be solved by numerical methods only. Since this system of equations includes all nonlinear hydrodynamic terms, we call the intermediate model by the refined model, in comparison with the simplest model presented below, which takes into account only the largest terms of the equations. The system (24)-(26) can determine the quantities $\psi, v_\varphi, \rho$ at the stage of intensification of the structure.

Finally, within the framework of ideal hydrodynamics, one can obtain one separate equation for the stream function, which describes either evolution of nonlinear internal gravity waves for $\omega_g^2 > 0$, or generation of the structure at initial stage for $\omega_g^2 < 0$. According to (Onishchenko, Pokhotelov, Horton, 2015), such an equation can be derived from (24) and (26) with the following simplifications (see expression (1)):

$$|\tilde{\rho}| \ll \rho_0, |\tilde{p}| \ll p_0, \frac{\tilde{p}}{p_0} < \frac{\tilde{\rho}}{\rho_0}, \tilde{\chi} \sim \psi, v_z \gg v_r, \frac{\partial}{\partial r} \gg \frac{\partial}{\partial z},$$

$$L < H, \frac{\partial}{\partial r} \gg \frac{1}{L_\rho}, L_\rho^{-1} = \frac{d\, ln\, \rho}{dz}, \frac{\omega^2}{c_s^2 k_r^2} L_\rho \frac{\partial}{\partial z} \ll 1,$$

here the last inequality assumes low frequency processes.

As a result, we have the following equation for the stream function:

$$\frac{\partial^2 \Delta_r^* \psi}{\partial t^2} + \omega_g^2 \Delta_r^* \psi + \frac{1}{r}\frac{\partial}{\partial t} J(\psi, \Delta_r^* \psi) = 0, \tag{27}$$

where $\Delta_r^*$ is the radial part of the operator (22). The Jacobian determinant represents the so-called vector nonlinearity. The simplest equation (27) allows us to find the stream function $\psi$ and, therefore, the radial and vertical components of the plasma velocity. The change in density is given by equation (24), and the azimuthal component of velocity is given by equation (25). This system of equations (24), (25), (27) describes the simplest model and can determine the quantities $\psi, v_\varphi, \rho$ at the stage of origination of a dynamic plasma structure. It should be emphasized that all models also take into account the pressure gradient (this variable is simply excluded using mathematical transformations).

Next, we will discuss the solutions that follow from the obtained systems of equations for all three cases, starting with the simplest one. Since for the simplest model with a cylindrical plasma structure a purely analytical solution can be obtained, to compare all three models we will first consider cylindrical plasma structures. The case of an expanding conical plasma flow as applied to jets can be solved numerically; we will consider such a case later within the framework of the full model in Section 7. There we will also discuss in detail the downstream flows in the solution used.

## 4 The velocity field in the framework of the simplest model

We will consider the convectively unstable case with $-\omega_g^2 \equiv \gamma^2 > 0$ (let us remember that the reasons for instability can be different, but the description of the phenomenon will be identical). Equation (27) can easily be solved by the method of separation of variables. To do this, we note that the sum of the first two terms can be zeroed out due to a time dependence of $\psi$, and the last term can be vanished due to some specific radial dependence of $\psi$. The algorithm for finding analytical solutions is described in detail in (Onishchenko, Artekha, Artekha, 2024). Here we will present only the final formulas (as the Bessel functions) needed for calculations and comparison with the other two models, and also give some new additional comments. We will use the following solutions:

$$\psi^{int}(t,r,z) = v_0 r r_0 f\left(\frac{z}{L}\right)\sinh(\gamma t)\, J_1(\delta_0 r/r_0), \tag{28}$$

$$\psi^{ext}(t,r,z) = v_0 r r_0 f\left(\frac{z}{L}\right)\sinh(\gamma t)\, K_1(\delta_0 r/r_0), \tag{29}$$

$$v_r^{int} = -v_0 \frac{r_0}{L} f'(Z)\sinh(\gamma t) J_1(\delta_0 r/r_0), \tag{30}$$

$$v_r^{ext} = -v_0 \frac{r_0}{L} f'(Z)\sinh(\gamma t)\, m K_1(\delta r/r_0), \tag{31}$$

$$v_z^{int} = v_0 f(z/L)\sinh(\gamma t)\, \delta_0 J_0(\delta_0 r/r_0), \tag{32}$$

$$v_z^{ext} = -v_0 f(z/L)\sinh(\gamma t)\, m\delta K_0(\delta r/r_0), \tag{33}$$

where $v_0 = const$ is some characteristic poloidal velocity ($v_0$ is introduced only for the simplest model), $L = const$ is the characteristic vertical spatial scale (the height of a plasma structure), $r_0$ is the characteristic radius of the structure. The characteristic values of $r_0$ and $L$ can vary greatly for different plasma structures (in further, only for some definiteness, in numerical estimates we will consider jets with $L = 100000$ km). Constants $\delta$, $\delta_0$ determine the profile of the velocity field of the structure, and $m$ ensures the smoothness of the function for all $r$. The function $f(Z)$ of the dimensionless parameter $Z = z/L$ determines the dependence of radial and vertical velocity components on height (boundary values, location of maxima etc.). For the simplest model, the values of this function at different $z/L$ act as free parameters. In this case, it is physically limited only by the conditions of the finiteness of the plasma structure itself, i.e. there is significant arbitrariness in the choice of the vertical structure. We can compose a single continuous solution from the piecewise exact solutions found above, as from blocks (by $r$ and by $z$), so that it satisfies all the physical conditions on the axis and on the periphery of the structure.

For the linear function $f(z/L)$, the azimuthal component of velocity can be obtained in quadrature for each area $i$:

$$v_\varphi = \exp\{c_0(\cosh(\gamma t) - 1)\}\, f_0(z/L) V_{\varphi r}(R), \tag{34}$$

$$V_{\varphi r}^i(R) = C^i \exp\left\{\int_1^R \frac{-\alpha^i x - n x \hat{V}_z^i(x) - \hat{V}_r^i(x)}{x \hat{V}_r^i(x)}\, dx\right\}, \tag{35}$$

where $\hat{V}_r^i(R) = \tilde{v}_r^i/r_0, \hat{V}_z^i(R) = \tilde{v}_z^i/L$, and velocity components with the tilde icon mean the factor of the same functions (30)–(33), but without temporary dependence and $z$-dependence. For example, for two regions in $z$, we write:

$$f(Z) = \begin{cases} k_1 Z, & 0 \le Z \le Z_1; \\ \frac{k_1 Z_1}{1-Z_1}(1-Z), & Z_1 < Z \le 1, \end{cases} \quad (36)$$

$$f_0(Z) = \begin{cases} k_2 Z^n, & 0 \le Z \le Z_1; \\ \frac{k_2 Z_1^n}{(1-Z_1)^n}(1-Z)^n, & Z_1 < Z \le 1, \end{cases} \quad (37)$$

where $Z_1$ is the height where the maximum is reached, $n > 0$ – any positive real number.

Within the framework of this model, we can describe both convective plasma structures with velocities $v_z \sim 7 \div 15$ km s$^{-1}$, and the initial stage of jet generation, which is what we are especially interested in. As an example, we will use the constants:

$$\delta_0 = 1.8412, r_1 = 1.7377, \delta = 2.5, \mathrm{m} = 30.9971, n = 0.8, c_0 = 0.01, \quad (38)$$

$$v_0 = 10 \text{ km s}^{-1}, \ r_0 = 0.02L. \quad (39)$$

We can construct a dynamical plasma structure from more than two linear parts in $z$. For the function $f(z/L)$, we will select 8 linear sections in $z/L$: (0, 0.05), (0.05, 0.1), (0.1, 0.2), (0.2, 0.45), (0.45, 0.55), (0.55, 0.75), (0.75, 0.9), (0.9, 1), with corresponding slope tangents: 9, 6, 1.5, 0.4, -0.1, -1.25, -2.4, -3.8. As a result, for the initial stage of jet generation (extrapolated to $\gamma t =$ 3.6), we obtained the velocity distribution in Fig. 1a. Formally, it is possible to achieve arbitrary observed jet velocities, but the range of applicability of the model is limited to velocities $v_z < c_s$. Note that we specifically choose such parameters so that all flows are distinguishable (in fact, all flow values and their localizations can vary within very wide limits, which will be discussed further in Section 7). On the $r - z$ plane, the first graph depicts the radial velocity component, the second graph shows the vertical component, and the third graph depicts the azimuthal velocity component. We see the spatial localization of flows inside a jet with extrema at certain distances $r$ and $z$; all movements disappear at the periphery of the structure. In this case, the velocity field and the order of all quantities resembles what is observed for the real solar jets (Raouafi et al., 2016; Shibata and Magara, 2011; Tian et al., 2021).

For the simplest model, the maximum speed of the jet is determined by the free parameter $v_0$ and the structure development time $t$. The speeds observed in the surrounding plasma region are chosen as the initial value of the speed $v_0$; they are significantly less than the speeds that will be achieved in the plasma structure under consideration after its self-organization. However, the simplest model is intended to describe only the initial stage of jet generation. Of course, the presented results are an extrapolation for the jet development time, since the simplest model does not take into account dissipative processes. Note that if we take other values of the parameters $L, v_0, t, \gamma$, then we can obtain velocity fields characteristic of other solar structures (granules, some spicules, etc.).

If we choose some nonlinear function $f(z/L)$, the dependencies of azimuthal velocity on variables $z$ and $r$ in equation (25) cannot be separated. Therefore, this equation should be solved numerically. However, in many cases the solution quickly becomes turbulent (with rotations in mutually opposite directions). For definiteness, we choose the following function:

$$f(z/L) = \tanh\{20z(L-z)/L^2\}. \qquad (40)$$

Then we obtained the velocity field shown in Fig. 1b (we chose $\gamma t = 3.6$ in the simplest model for extrapolation). As we see in Fig. 1b, many regions with multidirectional azimuthal velocities arise. Taking the discarded terms (magnetic, viscosity etc.) into account could lead to a preferential direction of rotation of such a turbulent structure. Perhaps, in such cases, it is necessary to search for a numerical solution of equation (25) simultaneously for all variables, including time. Again, different parameter values may describe different solar plasma structures.

As we see from the structure of equation (27), arbitrary dependence on $z$ can be included in the Brunt–Väisälä frequency $\omega_g(z)$, and the solution for the stream function $\psi$ will remain the same. There arises some dependence of growth rate on height $z$: $\gamma \to \gamma(Z)$. The same change will occur for the expression of the vertical velocity component, but for the radial velocity component the following additional replacement will occur: $f'(Z) \to f'(Z) + f(Z)\gamma'(Z)t\coth[\gamma(Z)t]$. Again, the equation (25) must be solved numerically.

The range of applicability of the simplest model is limited by the region of convective instability and the region with non-zero torque. The analytical solution can describe only the initial stage of jet generation (the simplest model does not consider dissipative processes).

Let us now pass on from the simplest convective model to the study of the refined model that takes into account all nonlinear terms within the framework of ideal hydrodynamics.

## 5 The velocity field of plasma structures in the framework of the refined convective model

Now let us consider the system of equation (24)-(26). In essence, the process in this case is rotating convection from the very beginning. Such a system can only be solved numerically by specifying the initial and boundary conditions. Let us recall that in ideal hydrodynamics we cannot set no-slip conditions for the tangential components of velocity at the boundary, but only non-passage conditions for the normal component of velocity. However, there are no solid surfaces in the Sun's atmosphere, so some boundary conditions can be quite arbitrary, except at $r = 0$ we must have $v_r = 0, v_\varphi = 0$. Note that the function $f(z/L)$ and the quantity $v_0$ are introduced only for the simplest model. However, for the refined model such parameters are not introduced, and integration is performed taking into account the boundary conditions. It is the boundary conditions that determine the vertical structure of the jet. The boundary conditions can be completely different, but we will choose the similar conditions (as in the solution for the simplest model) only to compare the results of all three models. Therefore, let's choose parameters (38), (39) and (40).

At the initial stage, the influence of the azimuthal velocity component on the profile and the magnitude of the radial and vertical velocity components is rather small. However, the azimuthal velocity component itself depends significantly on the initial and boundary conditions (there arises some unidirectional differential rotation, but it can be small or local). We can leave the corresponding solutions for the simplest model as initial and boundary conditions. In this case, the entire column $L$ is involved in the rotation, as is shown in Fig. 1c (here we chose extrapolation to $\gamma t = 3$). For the solar chromosphere and corona in the region of a plasma structure, we use the following approximation:

$$\rho_0(z) = \rho_{00} \frac{T}{T + l_0 z}, \tag{41}$$

where $l_o = 1 \div 10\ \mathrm{K\,km^{-1}}$, $\rho_{00} = 10^{-8} \div 10^{-9}\ \mathrm{kg\,m^{-3}}$, $T = 15000 \div 50000\ \mathrm{K}$. We used $l_o = 10\ \mathrm{K\,km^{-1}}$, $\rho_{00} = 10^{-8}\ \mathrm{kg\,m^{-3}}$, $T = 20000\ \mathrm{K}$. Thus, some of the solar spicules and jets (excluding fast coronal jets) can be explained by a convective mechanism (or the release of thermal energy). Note also that when performing numerical calculations in this model, we can calculate plasma structures not only in the shape of a cylinder, but conical and more complex shapes with cylindrical symmetry. Recall (Onishchenko, Artekha, Artekha, 2024) that the convective mechanism works well in the case of a purely vertical magnetic field (for example, near the poles) or with small magnetic field gradients on the jet scale. For both the simplest model and the refined model, we did not write out expressions for the generated magnetic field. In the case of ideal MHD, it can be found separately from the equation for the magnetic field to be frozen into the plasma:

$$\frac{\partial \mathbf{B}}{\partial t} = \nabla \times (\mathbf{v} \times \mathbf{B}),$$

or you can use Chandrasekhar's solution (see, for example, (Onishchenko and Feygin, 2022)). In the next Section we will plot the magnetic field for the full model, where all the equations are related.

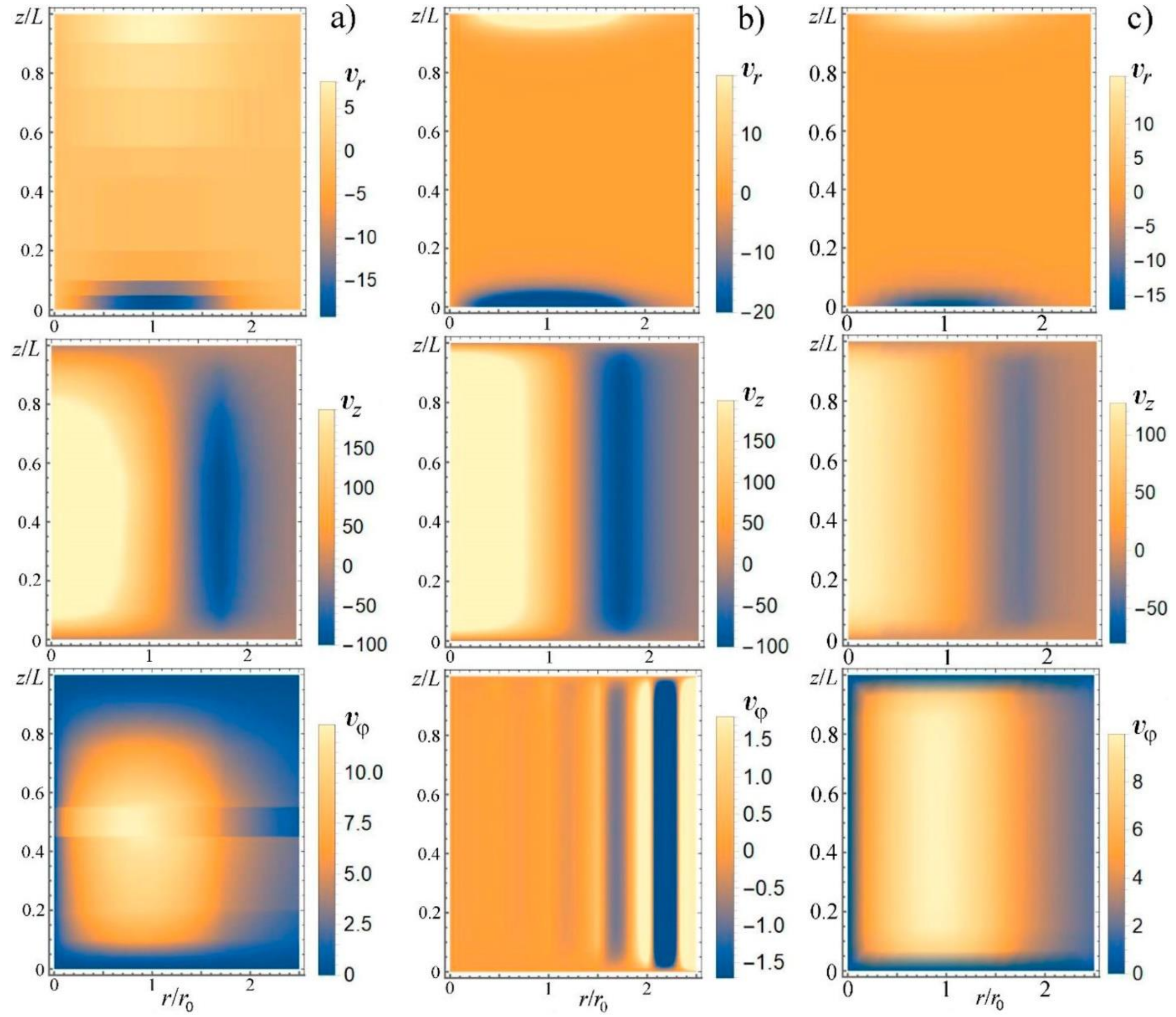

**Fig. 1** The velocity field of a dynamical plasma structure in $\mathrm{km\,s^{-1}}$ depending on the dimensionless coordinates $r/r_0$ and $z/L$ for the following hydrodynamic models: a) the simplest model for the linear profiles of the function $f(z/L)$; b) the simplest model for the nonlinear function $f(z/L)$ from (40); c) the refined model

Let us recall that in the simplest model (Onishchenko, Artekha, Artekha, 2024) the growth of the radial and vertical velocity components was approximately exponential, and the azimuthal component grew according to super exponential law (slowly at first and then extremely fast). For the refined model, all velocity components grow more consistently. Since both of above considered models do not take into account dissipative processes, only the order of values and the structure of movements are subject to comparison with reality (and between the models). These results are again an extrapolation to development time of the plasma structure (Only the inclusion of dissipative processes in the full model limits the maximum speed of the jet).

Let us now move on from convective models to studying the complete system of equations (including influence of magnetic field and viscosity).

## 6 Model of plasma structures with the complete system of equations

Now let us consider the complete system of equations (14), (17), (19), (20) and (23). The full system of equations can only be solved numerically by specifying all the initial and boundary conditions. In the full model, we choose boundary velocities of the order of those average values of velocities that are constantly observed in the surrounding plasma region, and then calculate what velocities are achieved if a certain plasma structure begins to develop (the observed velocities of plasma structures vary within certain limits; in particular, the velocities of jets are distributed over very wide limits). Since the solar atmosphere does not contact with a rigid surface, the only definite boundary conditions will be $v_r = 0, v_\varphi = 0$ at $r = 0$; other conditions can be rather arbitrary. For example, we can choose the boundary conditions as for an impenetrable (rigid) surface or choose the solutions of the simplest model. Let us recall that the function $f(z/L)$ and the quantity $v_0$ are introduced only for the simplest model; for the full model such parameters are not introduced, and integration is performed taking into account the boundary conditions. It is the boundary conditions that determine the vertical structure of the jet. The boundary conditions can be completely different, and we can choose the solutions of the simplest model (only to compare the results of all three models) with all the previous parameters as remaining boundary conditions for $v_r, v_\varphi, v_z$. In the latter case, the conditions can be chosen to be either constant or changing over time: the solution will adapt to these changes. We will apply expression (41) as the initial equilibrium density.

For the hydrogen plasma, we can evaluate specific conductivity and dynamic viscosity from Spitzer's formula and Braginskii's formula, respectively:

$$\sigma = \frac{3\varepsilon_0^2 (2\pi kT)^{3/2}}{e^2 \sqrt{\pi m_e} ln\Lambda}, \ \eta = 1.2 \cdot 10^{-17} \left(\frac{20}{ln\Lambda}\right) T^{5/2},$$

where the vacuum permittivity is $\varepsilon_0 = 8.85 \cdot 10^{-12}$ F m$^{-1}$, the Boltzmann constant is $k = 1.38 \cdot 10^{-23}$ J K$^{-1}$, electron mass is $m_e = 9.11 \cdot 10^{-31}$ kg, electron charge is $e = 1.6 \cdot 10^{-19}$ C, the Coulomb logarithm is $ln\Lambda \sim 15$. Although the chromosphere and corona are generally stable according to the Schwarzschild criterion, if hotter plasma appears in a certain area (positive temperature gradient arises), then the buoyancy mechanism is activated (local instability arises). The full model takes into account such effects, including pressure gradient. For definiteness, we will consider solar coronal jets, excluding fast jets (similarly, we can also calculate other solar structures). Although the orders of molecular constants (kinematic viscosity and specific conductivity) for the solar atmosphere at the bottom of the jet are: $\nu \sim 4 \cdot 10^3 \ \mathrm{m^2 s^{-1}}, \sigma \sim 4 \cdot 10^4 \ \mathrm{S \cdot m^{-1}}$, the typical dimensions of the structures are hundreds of kilometers. Therefore, we will use subgrid scale in the method of large eddy simulation (with the appropriate coefficients for these values).

Local magnetic fields on the Sun vary over a very wide range, and their approximations can be different for different cases. Moreover, direct measurements of the coronal magnetic field remain a challenging experimental task (most estimates are based on extrapolation of photospheric data, models, or indirect methods). There is no complete, universal analytical model that accurately describes all aspects of the magnetic field of such complex, dynamic and often non-stationary structures as jets and spicules. Currently, many extrapolations of the magnetic field have been proposed (Sakurai, 1989; Aschwanden, 2005; Judge, Tritschler, Low, 2011; Reale, 2014; Pariat et al., 2015; Raouafi et al., 2016; Martínez-Sykora et al., 2017; Wiegelmann and Sakurai, 2021). We will approximate the ambient magnetic field in the jet region by the following dependencies:

$$B_{0z} = B_0 f_b(t) \left(\frac{r_0}{r_0 + r + k_b z}\right)^2,$$

$$B_{0r} = B_0 f_b(t) \frac{k_b r r_0^2}{(r_0 + k_b z)(r + r_0 + k_b z)^2},$$

$$B_{0\varphi} = B_{02} f_b(t) \frac{v_{0z} r}{v_0 r_0}, \tag{42}$$

with some constants $B_{02} \sim B_0 = 0.002 \ \mathrm{T}, k_b = 0.0001$. We will use either constant function, or slowly doubling function $f_b(t) = 1 + \mathrm{Tanh}(\gamma t)$. For the axisymmetric case, we used the model of flux tubes, which relies on the conservation of magnetic flux across the cross-section of the tube. In this case $B_z = B_0 r_0^2 / r(z)^2$. We choose the most natural linear expansion of the magnetic tube at these scales, which corresponds to the first formula in (42). If a time coefficient is used, it is chosen so that the value changes smoothly from one value to another. From the strict equation $\mathrm{div}\, \mathbf{B}_0 = 0$ in cylindrical coordinates, the second expression in (42) strictly follows. The azimuthal component of the magnetic field depends mainly on the distributed vertical currents in the elongated magnetic tube. From Ampere's law we can obtain $B_\varphi \sim j_{0z} r \sim v_{0z} r$, which is used for the third formula in (42). Thus, the expression (42) is one of the possible axisymmetric analytical approximations: how the magnetic field in a given region of the Sun can change in terms of $r$ and $z$, taking into account the characteristic parameters of the plasma and their changes in the selected region. We can choose any other dependency; for the full model, this is a technical issue.

We numerically solve the 2D time-dependent, resistive MHD equations with spatial dependent viscosity and specific conductivity, in cylindrical geometry. The temperature stratification of real solar atmosphere will be approximated by a piecewise linear function (with constant temperature gradients). Linear temperature dependence leads to a power-law approximation of the density, which, for the chosen scales, coincides with expression (41) to within 0.1%. This dependence is used in computations. The initial conditions correspond to some background values of the quantities in the solar atmosphere. Computations are performed using the methods "Time Integration", "Explicit Euler" on a static grid (with subgrid scales) over the region $L = 0 \div 100000$ km, $R = 0 \div 5000$ km, with the working precision of 10 (the number of decimal places at the stage of intermediate calculations), the precision and accuracy goals of 4 (how many effective significant digits of accuracy should be achieved in the final result). These values are chosen to keep the computation time reasonable. To check the convergence of the numerical solution, the number of spatial grid nodes (with a corresponding change in subgrid scales) was doubled four times, starting from 20 and up to 320. By successively dividing the grid in half and running the system of equations multiple times (run-through method), a solution with four exact significant digits is obtained (improvement is carried out until this value is reached). As a result, the difference between the solutions decreased, and the last two calculations yielded similar results.

At $t = 600$ s and almost constant boundary conditions (impenetrable surfaces), the solutions for the velocity field is shown in Fig. 2a. As we found, all velocity components increase in a coordinated manner (in contrast to the simplest model, where the poloidal velocity components grew exponentially, but the azimuthal velocity component grew more quickly). The unidirectional differential rotation occurs for the entire column $L$. The vertical velocity component (upward flow) reaches a maximum value near the center of the jet core.

Our goal was to compare solutions at the same time moment, but with different values of the problem parameters. As a result, it was discovered that even an increased value of the ambient vertical magnetic field $B_0$ – up to 0.1 T – does not have a noticeable effect on the jet velocity field (regardless of whether the boundary conditions were fixed or the values at the boundaries smoothly increased). Furthermore, even an additional increase in the vertical gradient of the magnetic field (determined by the value of $k_b$ to $k_b = 0.001$) also does not have a noticeable effect on the jet velocity field. This numerical result confirms the theoretical analysis of this paper and the preliminary evaluation of the work (Onishchenko, Artekha, Artekha, 2024). Since locally (in $z$) the observed structure of jets is somewhat closer to axisymmetric (as in our models) than to flattened, it can be concluded that the ambient stationary magnetic field and its gradients in the volume of the jet do not have a determining effect on the structure of the jet and its velocity field. Apparently, only a rapid change in the non-uniform magnetic field with reconnection of the field lines has a significant effect on the process of jet generation.

The generated magnetic field itself adapts to the structure of the plasma flows and the boundary conditions; for example, the radial component of the magnetic field can become comparable with other field components with a large gradient of the surrounding magnetic field (with a large value of $k_b = 0.01$). However, we are more interested not in the generated magnetic field, but in the structure of plasma flows and the velocity field.

Note that some change in surrounding temperature or temperature gradient in the equilibrium density (41) chosen here also has little effect on the velocity field, but a super adiabatic temperature gradient at the bottom of the jet is needed for instability itself (we assume that the excess energy arose at $z < 0$).

In our case, the solutions for the velocity field in the volume of the jet are adjusted to the conditions at the boundaries. The jet velocity field begins to increase noticeably as soon as we set non-zero fluxes at the boundaries (both for stationary boundary conditions and for those increasing with time). We set the boundary flow velocities ($v < 1$ km s$^{-1}$) much lower than the characteristic plasma velocities. The contour plot for the jet velocity field in Fig. 2b depicts such a case, where $t =$ 900 s. Computations show that all velocity components increase in a coordinated manner. The generated magnetic field for the large gradient of the surrounding magnetic field (with $k_b = 0.01$) is shown in Fig. 2c. Note that for other initial and boundary conditions, we can obtain velocity and magnetic fields characteristic of different solar structures (granules, some spicules, some jets, etc.).

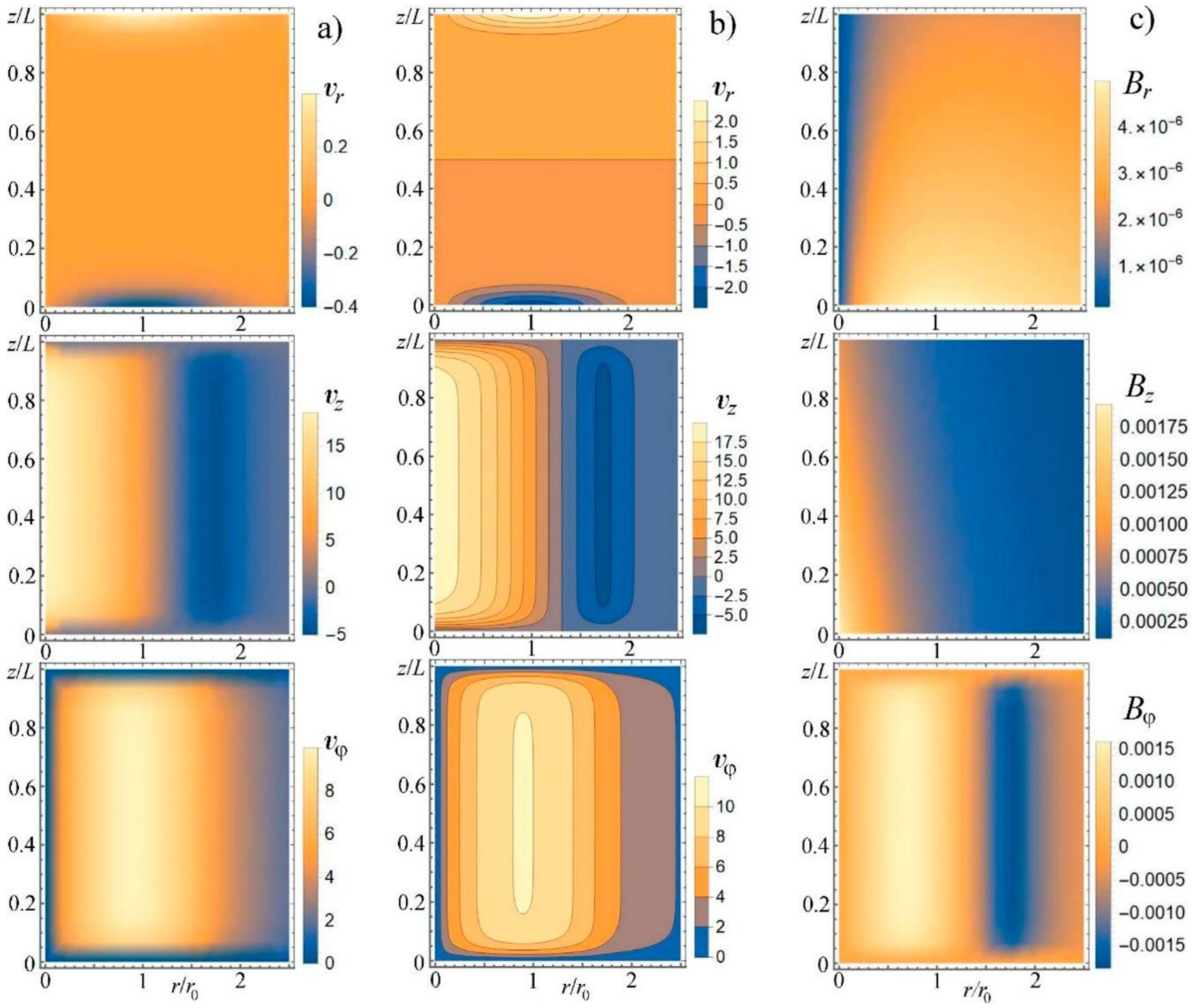

**Fig. 2** Computation results for the full model of a dynamical plasma structure. The following quantities are presented as functions of the dimensionless coordinates $r/r_0$ and $z/L$: a) the velocity field in $\mathrm{km\ s^{-1}}$; b) the contour plot for the velocity field in $\mathrm{km\ s^{-1}}$ in the presence of non-zero flows at the boundaries; c) the generated magnetic field in T

## 7 Discussion of results and conclusions

A possible mechanism of generation of some dynamic plasma structures can be associated with the Schwarzschild criterion for convective instability (or with the release of excess of any energy). Due to the super adiabatic temperature gradient for some plasma layer, the overheated plasma mass starts to rise. Since the motion obeys the continuity equation, excess thermal energy leads to increased vertical flow and intensification of the vortex motion.

It is well known that within the convective zone of the Sun, the convection process is constantly present. However, convective processes are also observed in the solar atmosphere, since the temperature continues to decrease with increasing altitude throughout the photosphere (where granules are observed) and in the lower part of the chromosphere, down to the minimum temperature $T = 4400$ K. Therefore, the proposed models can be applied to solar granules (this is the simplest and most natural application). In the very lower part of the chromosphere, conditions for the formation of some spicules can also be realized. Let us mention a few statements: “Photospheric velocities are dominated by convective granulation” (De Pontieu, Erdélyi, James, 2004); “both types of spicules are tied to the relentless magnetoconvective driving in the photosphere” (De Pontieu et al., 2007). Pike & Mason (1998) firstly described “solar tornadoes”, which connect the convection zone to the upper solar atmosphere (it is known that tornadoes are caused by a thermal mechanism). They used this term to describe macrospicules seen by the Solar and Heliospheric Observatory (SOHO). Thus, the thermal mechanism plays a certain role in these phenomena, and it was necessary to clarify this role. The generally accepted view on the formation of solar jets is that magnetic field reconnection plays a key role, but even the authors of the review (Raouafi et al. 2016) point out that despite significant progress, there are many unclear issues, and the physical mechanisms are not fully understood. Inferring magnetic reconnection based solely on observations is not straightforward (Li et al., 2025). In the work (Chandrashekhar et al., 2014), where HMI magnetograms were used, no flux emergence was detected in the jetting region. The “anemone” model, in which the jet is the result of a dipolar magnetic structure reconnecting, exhibits shortcomings in explaining such properties as helical structures and apparent transverse motions (Raouafi et al., 2016, P. 28). A single violent burst reconnection event seems unable to explain the very long lifetime of the observed jet, its constant drift motion at different altitudes, and its constant transverse speed (Chandrasekhar et al., 2014). It turned out that the speed of ascent was much lower than the expected Alfvén speed. Since most measurements of jet apparent speeds are around 200 km s−1, which is smaller than sound speed, then it is concluded in (Shimojo and Shibata, 2000) that chromospheric evaporation can be considered a strong candidate for jet acceleration. Let us recall that the evaporation mechanism consists of the fact that superheated plasma (i.e. the thermal mechanism is also important here) compressed by a magnetic field is released and accelerated along the magnetic field lines (Fang, Fan, McIntosh, 2014). In active regions, the number of thermally-driven jets is larger than the magnetically-driven ones (Sako, 2014). For an EUV jet which occurred in an active region, the cool plasma (observed in He II and Fe VIII lines) was ejected by magnetic acceleration due to magnetic reconnection, while the hot plasma (observed in Fe XII and Fe XVI lines) was ejected by thermal acceleration similar to the chromospheric evaporation (Matsui et al., 2012). Thus, there are many examples of the influence of the thermal mechanism on the development process of the plasma structures under consideration. We do not claim that any type of jet or spicule can be formed solely due to convective instability (convection is not typical for other regions of the Sun). We only suggest that the Schwarzschild instability and the thermal mechanism can also play an important role in the process of plasma structure formation. As some confirmation, we would like to draw attention to the fact that the dynamic structures considered in the article have a temperature higher than the typical temperature of those regions of the Sun from which they began to develop and spread upward. Since the lower part of the future plasma structure was initially overheated (there observed a bright point), the preceding rise of this region of overheated plasma (some "plasma bubble") could create a local region with non-stationary stratification. As a result, the Schwarzschild instability mechanism may be triggered locally (along with other possible mechanisms).

In this work, a complete system of equations is derived, which includes an equation for the stream function, an equation for the azimuthal component of the velocity, an equation for the change in density, an equation for the flow function for poloidal magnetic field, and an equation for the azimuthal component of the magnetic field. The complete system is applicable for all stages of development of dynamic plasma structures in the viscous case (this is the full model). Some simplification of the system of equations for the case of ideal hydrodynamics was also carried out (but still considering a rotating case). This system takes into account all nonlinear hydrodynamic terms and is applicable for the growth stage of the structure (this is the refined model). Finally, a separate equation for the stream function is obtained, applicable to the initial stage of vortex generation (this is the simplest model). This is one of the possible methods for describing convective instability. All presented models describe the generation and growth (intensification) of axially symmetric structures.

Within the framework of the simplest model, an analytical localized solution can be obtained. Some initial twist plays an important role in the vortex generation: it symmetrizes poloidal movements about the axis and breaks the symmetry with respect to the direction of rotation. The reasons for the initial rotation can be different. Rotation with some definite profile increases due to the conservation of angular momentum (the law of conservation of mass and angular momentum is satisfied in this case). Due to the radial inflow, the plasma structure will intensify in the instability region with non-zero twist (non-zero helicity). The analytical simplest model is quite flexible and allows us to describe a wide variety of structures: spatial structure, structure of the velocity field, values of extreme velocity components, their location etc. The resulting velocity field is shown in Fig. 1a. In the lower part of the plasma structure, the radial flows converge to the axis (the maximal inflow near $r \sim r_0$). In the upper part of the plasma structure, inflows and outflows change places. The upward flow is maximal at the center of the axis of the plasma structure. Plasma rotation is differential. In the case of a nonlinear z-profile of the vertical flow, many regions with opposite rotation directions arise (see results in Fig. 1b).

Only in order to compare the obtained solutions with the analytical solution for the simplest model, all subsequent solutions are obtained for cylindrical plasma structures. Next, for the refined and full models, such parameters as $f(z/L)$ and $v_0$ are not introduced, and integration is performed taking into account the boundary conditions. It is the boundary conditions for these models that determine the vertical structure of the plasma object. The boundary conditions can be completely different (arbitrary). However, solely for the purpose of comparing the results of all three models, we have chosen boundary conditions that coincide with the solutions of the simplest model at these boundaries. The simplest and refined models are intended to describe the initial stage of development of the process, since they do not consider dissipative processes. Only the inclusion of dissipative processes in the full model limits the maximum speed of the jet. Within the framework of the refined model, some unidirectional differential rotation occurs (see Fig. 1c). However, the value of the azimuthal velocity component depends on the initial and boundary conditions.

The presented auxiliary hydrodynamic (simplest and refined) models can serve to preliminarily evaluate the possible structure of the velocity field solution and select appropriate conditions before resorting to complex and time-consuming numerical computations in the complete system. Any magnetic field always has effects through magnetic pressure and magnetic tension. Of course, magnetic field effects must be taken into consideration when modeling the formation of any jets and spicules (Raouafi et al., 2016; De Pontieu, Erdélyi, James, 2004).

In the full model, we choose boundary conditions of the order of those average values that are constantly observed in the surrounding plasma region and then calculate what velocities are achieved if a certain plasma structure arises. Within the framework of the full model, jet development is considered as rotating structure from the very beginning to the stage of greatest intensification. All velocity components increase in a coordinated manner. The excess energy in the viscous case can turn into kinetic energy of rotation and increase the angular momentum ("heat engine"). Unidirectional differential rotation occurs. The velocity component profiles are smoothed (see Fig. 2a). The ambient stationary magnetic field and its gradients in the volume of the jet do not have any noticeable direct effect on the structure of the jet and its velocity field. The velocity field inside the jet is adjusted to the conditions at the boundaries: the velocity field begins to increase as soon as we set non-zero fluxes at the boundaries (the corresponding contour plot is shown in Fig. 2b). The structure of generated magnetic field is shown in Fig. 2c. Really, the strong magnetic field always has the following effect on a jet: it collimates plasma flows (we choose the corresponding $r_0$) and creates a preferential direction of propagation ($v_z$) along the magnetic field. As a result, the vertical component of the jet velocity (in the case of a predominantly vertical magnetic field) turns out to be much greater than the radial and azimuthal components (this influence is taken into account through the initial and boundary conditions).

The computations carried out showed that all models of axisymmetric plasma structures (with cylindrical shape) yield results similar in structure and magnitude (taking into account some time extrapolations for the simplest and refined models). Therefore, the simplest model can serve as a test model, for preliminary selection of all conditions and verification of the expected results before performing very cumbersome and time-consuming numerical computations in the complete model. This is the main purpose of the simplest analytical model (and this was the point of comparing all three models).

However, for numerical calculations, the shape of the plasma structure can be chosen not only as cylindrical, but also, for example, as conical. As an example, we substitute $r \to r^2/(r + z/25)$ in the dependencies for the initial and boundary conditions (we again use the simplest model to set the values of all functions on the boundaries for subsequent computation of the full model); we chose the following function:

$$f(z/L) = \begin{cases} -0.05 - \tanh[12z\,(L-z)/L^2]\,, & z < 0.6L\,, \\ -0.55 - 0.513\tanh[11z(0.9L-z)/L^2]\,, & z > 0.6L\,. \end{cases} \tag{43}$$

The results of calculating the velocity field for the full model with rather slow boundary flows are shown in Fig. 3. We see that the radial outflow at the top of the plasma structure is almost imperceptible, and the downward surrounding flow is noticeably smaller than the upward flow along the axis.

It should be noted that we specifically chose such parameters of the simplest model (and such boundary conditions for the refined and full models) so that the flow structures were clearly distinguishable (it is for this reason only that the descending flows were located close to the main ascending flow along the radius and turned out to be rather noticeable). Convective flow fields, including downflows, are regularly observed in plasma with the high and intermediate beta parameter (for example, in the convective zone, photosphere and chromosphere of the Sun). Even in the solar corona, downward flows are a well-observed phenomenon associated with solar flare current sheets and open-closed magnetic flux boundaries. Although downward flows have not yet been detected by satellite measurements near coronal jets (especially jets in the open field region of coronal holes), it is clear that at jet speeds less than the orbital velocity (for the Sun $v_I \approx 437\ \text{km s}^{-1}$) some kind of return downward motion around must arise (see also (Chandrashekhar et al., 2014)). Simply, if an outflow from the axis and such a downward movement of plasma surrounding the jet will occur in a large volume ($\Delta r \gg r_0$), then its speed will be too small to be detected (the influx to the axis can also be collected from large $\Delta r \gg r_0$ in the lower plasma layer). Such a flow structure can be easily obtained if, instead of (38), other parameter values (in the case of the simplest model, and for other two cases, the corresponding boundary conditions) are selected. For example, we can choose the following new parameters:

$$r_1 = 1.3121, \delta = 0.08, \text{m} = 0.0551, L \gg 10r_0, v_z(0,0,0) < 2\ \text{km s}^{-1}, \quad (44)$$

the function (43), and small flows at the boundaries (conical shape). The results of calculating the velocity field for the full model are shown in Fig. 4.

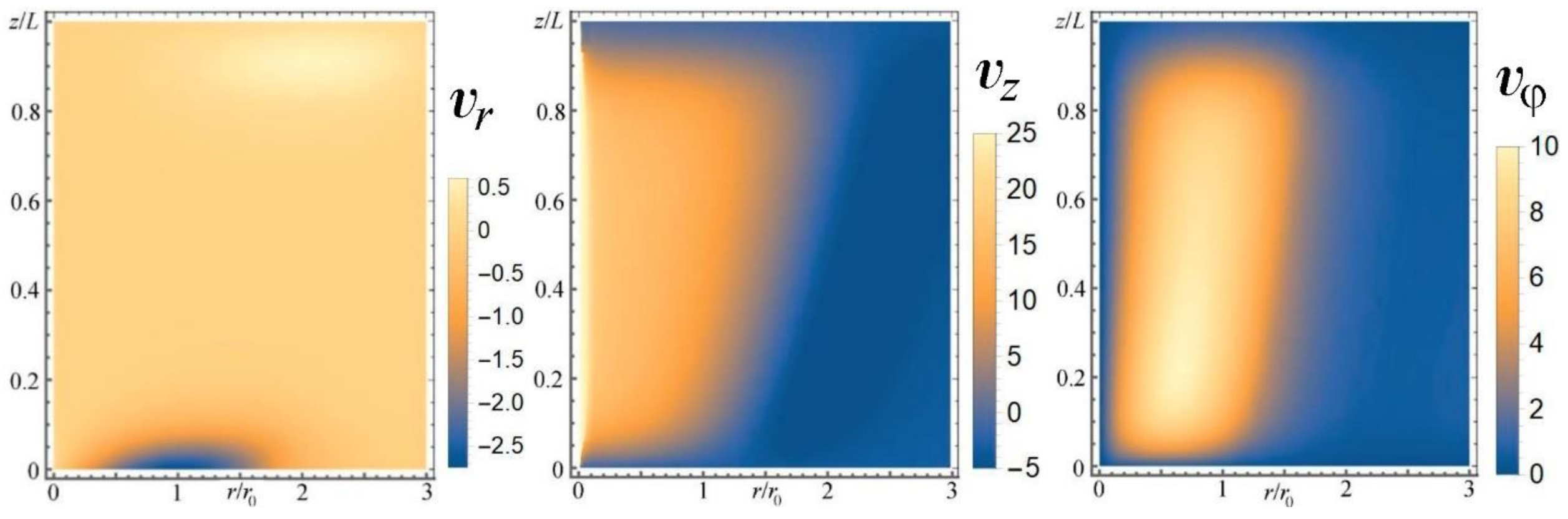

**Fig. 3** The velocity field of a plasma structure in km $\text{s}^{-1}$ depending on the coordinates $r/r_0$ and $z/L$ for the full model and the structure of conical shape

We see that the radial outflow at the top of the jet is insignificant and can only be detected by a slight radial expansion of the jet. The speed of the downward flow of the surrounding plasma is almost two orders of magnitude less than the speed of the upward flow of the jet itself, and therefore this movement is practically unnoticeable. The greater the angular momentum of the plasma layer, the greater the arising azimuthal velocity of the jet, which can become comparable to the vertical velocity of the jet itself. Compared to the previous plasma structures, Fig. 4 most closely resembles real solar jets.

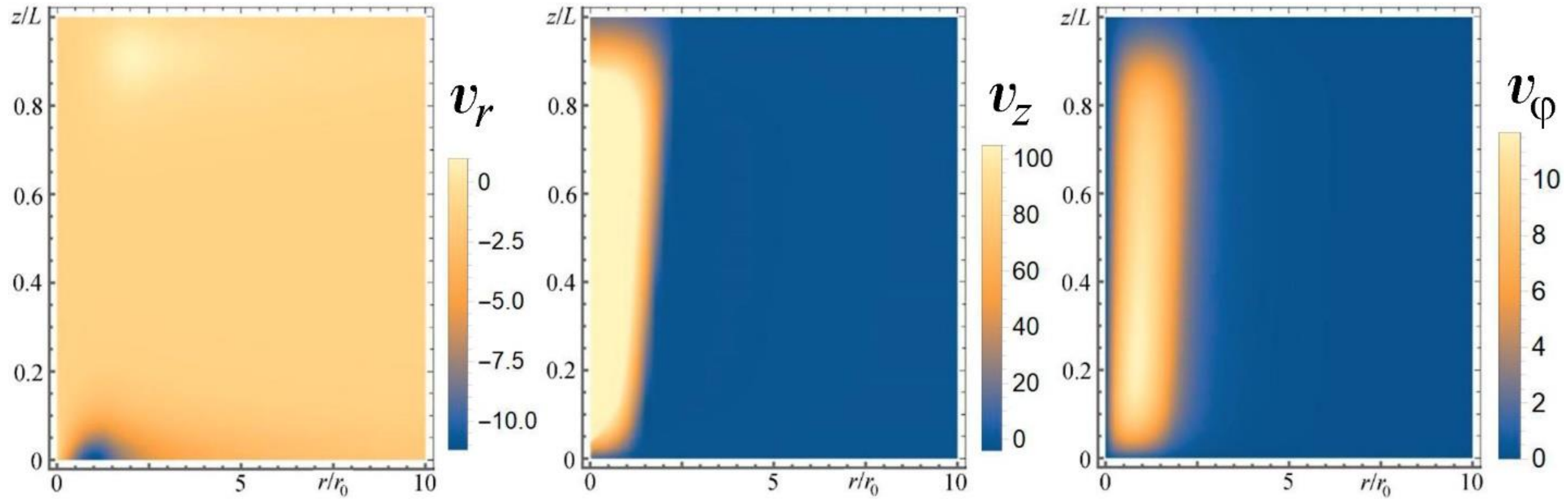

**Fig. 4** The jet velocity field in km $s^{-1}$ depending on the dimensionless coordinates $r/r_0$ and $z/L$ for the full model and jet structure of conical shape. Here the outflow from the axis and the downward flow are practically imperceptible

The presented solutions can describe models for certain types of photospheric, chromospheric and convective flows, including magnetic field generation. To compare the obtained numerical solutions of the full model with observations, we present some characteristics of the surrounding plasma and the corresponding dynamic plasma structures in the solar atmosphere: granules (Bray, Loughhead, Durrant, 1984; Kostyk and Shchukina, 2004; Puschmann et al., 2005; Kostik, Khomenko, Shchukina, 2009; Malherbe et al., 2015; Castillo et al., 2022), spicules (Sterling, 2000; De Pontieu, Erdélyi, James, 2004; De Pontieu et al., 2007; Judge, Tritschler, Low, 2011; Tsiropoula et al., 2012; Martínez-Sykora et al., 2017), jets (Scullion et al., 2009; Wedemeyer-Böhm et al., 2012; Matsui et al., 2012; Chandrashekhar et al., 2014; Cheung et al., 2015; Raouafi et al., 2016). The main characteristics of the solar atmosphere (there are some differences depending on the solar activity and different sources and models: Bilderberg, HSRA, VAL, HOLMU, UNLV, Waldmeier) are summarized in Table 1.

**Table 1** Some averaged characteristics of the solar atmosphere. *The minimum temperature of the chromosphere reaches 4400 K at an altitude of 100 km

| | Photosphere | Chromosphere | Transition region | Corona |
|---|---|---|---|---|
| Height, km | 400 | 2500 | 200 | $\sim 10^7$ |
| $T_{bottom}$, K | 6400 | 4700* | 25000 | $10^6$ |
| $T_{top}$, K | 4700 | 25000 | $10^6$ | $1.5 \times 10^6$ (at $10^5$ km) |
| $\rho_{bottom}$, $\mathrm{kg \cdot m^{-3}}$ | $3 \times 10^{-4}$ | $3 \times 10^{-6}$ | $10^{-9}$ | $10^{-11}$ |
| $\rho_{top}$, $\mathrm{kg \cdot m^{-3}}$ | $3 \times 10^{-6}$ | $1 \times 10^{-9}$ | $10^{-11}$ | $2 \times 10^{-16}$ (at $10^5$ km) |

Using the data from Table 1, for the photosphere at zero boundary conditions (as on a solid surface) we obtain: $r_0 \sim 500$ km, $v_z \sim 2 \div 7$ $\mathrm{km \cdot s^{-1}}$, which corresponds to the characteristics of the granules (time of structure development $\tau \sim 5 \div 10$ min). Using the data for the lower part of the chromosphere, the boundary conditions $v_{z0} \sim 0.2 \div 0.4$ $\mathrm{km \cdot s^{-1}}$ as in (De Pontieu, Erdélyi, James, 2004) and $T \sim 8000$ K, we obtain $r_0 \sim 250$ km, $v_z \sim 20 \div 40$ $\mathrm{km \cdot s^{-1}}$ (time of structure development $\tau \sim 5$ min), which corresponds to type I spicules (De Pontieu et al., 2007); for $T \sim 20000$ K we obtain $v_z \sim 50 \div 100$ $\mathrm{km \cdot s^{-1}}$, which corresponds to type II spicules. The jet is generated and reaches the stage of greatest intensification within 10-60 minutes (the computation results for jets and the corresponding graphs are given earlier; the boundary quantities change from background values to some maximal values). These final values of the velocity components correspond to the quantities observed in some solar jets. The models considered are applicable only to the description of some axisymmetric plasma jets arising in a strong vertical ambient magnetic field, when $v_z < c_s$ (slow standard jets, Hα surges). Beyond the applicability of these models are fast jets (X-ray jets, many of EUV-jets), blow-out jets, as well as cases of twisted magnetic flux ropes, asymmetric structure of flows and fields, and highly inhomogeneous and non-stationary magnetic fields. Thus, each proposed model (in its area of applicability) can describe rather adequately the main characteristics of some plasma structures.

Note that all the dynamic plasma structures we are considering have an elevated temperature compared to the average temperature of the surrounding plasma layer. Therefore, thermal energy plays an important role in the generation and maintenance of these structures in any case. Note that even fast coronal jets (which we do not consider in this article) have greatly elevated temperatures; therefore, at a minimum, this heating must be explained, since the same reconnection mechanism cannot simultaneously explain the emergence of both the jet's enormous unidirectional translational velocity and the spread of velocities in all directions that determines the plasma temperature. Essentially, it is necessary to determine what comes first (primary): the "external" magnetic field (controlled and reconnected by an unknown entity), which acts on the plasma, or the moving superheated plasma, which must self-consistently carry out this very reconnection of the magnetic field to eject the already superheated mass? Thus, in this case, too, the thermal mechanism (due to superheated mass rising from the depths of the Sun) must at least play the role of a trigger (although we do not discuss fast coronal jets in the article).

It was previously shown how the proposed model explains the near undetectable nature of the reverse downward countercurrent. We can formulate some possible predictions of the proposed model for some spicules and jets, but it is necessary to take into account the fact that, as a rule, not one separate mechanism of the phenomenon is involved, and along with the thermal mechanism, other mechanisms can act simultaneously ($p$-modes and shock waves for spicules; reconnection of magnetic field lines for jets, etc.). Therefore, they are more in the nature of trends. At all other things being equal (mass, density, diameter), the higher the temperature of the plasma (from which the structure forms), the greater the vertical velocity can be achieved and the faster the plasma structure can reach its peak development (it should be noted that this is not the lifetime of the plasma structure - it is determined by the supply of thermal energy and instabilities). The temperature at the axis will be higher than at the side edges of the plasma structure. The probability of the formation of plasma structures with $D < D_{min}$ and $D > D_{max}$ drops sharply due to the development of instabilities.

The proposed model does not use artificial settings (pre-existing configurations that would themselves require an explanation within a self-consistent description of how they formed); it uses the most typical conditions of the surrounding plasma and takes viscosity and conductivity into account. Of course, this model also requires improvement (taking into account rapid non-stationarities, the possible influence of reconnection of magnetic field lines on movement and temperature, a more accurate accounting of the radial and vertical dependence of turbulent viscosity), which the author plans to do in the future.

In most of the model calculations, we assumed that the temperature gradient was constant throughout the entire height of the plasma structure. In fact, if a sudden emergence of a plasma mass from the depths of the Sun with a greatly elevated temperature occurs, the temperature gradient at the leading edge will be enormous. The movement of the plasma jet will resemble a directed explosion. As a modification, the model can be supplemented with a large temperature gradient with "running" maximum along the $z$-axis (the almost stepwise temperature distribution at the leading edge moves along the $z$-axis). This will result in a much higher maximum jet velocity $v_z$.

As a hypothesis, the increase in jet temperature and the super adiabatic temperature gradient can also be realized due to the Ranque–Hilsch vortex effect (the temperature separation) during the collision of swirling flows (Kalashnik and Visheratin, 2008) (apparently, this possibility has not been discussed previously). In the solar atmosphere, the simultaneous presence of vortex colder flows and of vortex hotter countercurrents indirectly confirms the manifestation of this effect.

The jet phenomenon is very complicated, so it is impossible to reveal all the connected nuances of this extensive research topic in one paper. Some important further direction in the development of the suggested jet models is the search for quantitative relationships of found parameters with the observed characteristics of an unstable plasma (including some statistical analysis of the fulfillment of the conditions for the formation of a jet).

**Acknowledgments** The work was carried out within the framework of the State assignment on the topic of fundamental scientific research "Plasma" IKI RAS.

**Author Contributions** S.N.A: Conceptualization, Writing – original draft, Visualization, Validation, Software, Methodology, Investigation, Formal analysis.

**Funding Information** No funding was received to assist with the preparation of this manuscript.

**Data Availability** No datasets were generated or analyzed during the current study.

**Declarations**

**Ethics declaration** Not applicable.

**Competing Interests** The authors declare no competing interests.